\documentclass[sigconf, nonacm]{acmart}

\usepackage{fancyhdr}
\fancypagestyle{firstpage}{
  \fancyhf{}
  
  \fancyhead[C]{\normalsize To appear in the 52nd IEEE/ACM International Symposium on Computer Architecture (ISCA), 2025.}
  \fancyfoot[C]{\thepage}
}

\AtBeginDocument{%
  }

\usepackage{algorithmic}
\usepackage{graphicx}
\usepackage{textcomp}
\usepackage{xcolor}
\usepackage{tikz}
\usepackage{graphicx}
\usepackage[normalem]{ulem}

\usepackage{pifont}
\newcommand{\cmark}{\textcolor{green}{\ding{51}}} 
\newcommand{\xmark}{\textcolor{red}{\ding{55}}}

\usepackage{xcolor}
\usepackage{xspace}
\definecolor{myBlue}{rgb}{0,0.3,0.5}
\newcommand{\revision}[1]{{\textcolor{myBlue}{#1}}}

\usepackage{threeparttable} 
\usepackage{caption} 
\begin{document}

\title{FlexNeRFer: A Multi-Dataflow, Adaptive Sparsity-Aware Accelerator for On-Device NeRF Rendering}



\author{Seock-Hwan Noh}
\affiliation{%
  \institution{DGIST}
  \city{Daegu}
  \country{Republic of Korea}}
\email{nosh3332@dgist.ac.kr}

\author{Banseok Shin$^\ast$}
\affiliation{%
  \institution{Samsung Electronics}
  \city{Suwon}
  \country{Republic of Korea}}
\email{shin.banseok@gmail.com}

\author{Jeik Choi$^\ast$}
\affiliation{%
  \institution{DEEPX}
  \city{Seongnam}
  \country{Republic of Korea}}
\email{cji@deepx.ai}

\author{Seungpyo Lee$^\ast$}
\affiliation{%
  \institution{Fitogether}
  \city{Seoul}
  \country{Republic of Korea}}
\email{lsp5490@gmail.com}

\author{Jaeha Kung$^\dagger$}
\affiliation{%
  \institution{Korea University}
  \city{Seoul}
  \country{Republic of Korea}}
\email{jhkung@korea.ac.kr}

\author{Yeseong Kim$^\dagger$}
\affiliation{%
  \institution{DGIST}
  \city{Seoul}
  \country{Republic of Korea}}
\email{yeseongkim@dgist.ac.kr}

\thanks{$^\ast$ Banseok Shin, Jeik Choi, and Seungpyo Lee were previously affiliated with DGIST.\\
$\dagger$ Jaeha Kung and Yeseong Kim are corresponding authors \{\textit{Email:} \textit{jhkung@korea.ac.kr} and \textit{yeseongkim@dgist.ac.kr}\}
}

\renewcommand{\shortauthors}{}

\begin{abstract}
Neural Radiance Fields (NeRF), an AI-driven approach for 3D view reconstruction, has demonstrated impressive performance, sparking active research across fields.
As a result, a range of advanced NeRF models has emerged, leading on-device applications to increasingly adopt NeRF for highly realistic scene reconstructions.
With the advent of diverse NeRF models, NeRF-based applications leverage a variety of NeRF frameworks, creating the need for hardware capable of efficiently supporting these models.
However, GPUs fail to meet the performance, power, and area (PPA) cost demanded by these on-device applications,  or are specialized for specific NeRF algorithms, resulting in lower efficiency when applied to other NeRF models.
To address this limitation, in this work, we introduce \textit{\uline{FlexNeRFer}}, an energy-efficient versatile NeRF accelerator.
The key components enabling the enhancement of FlexNeRFer include: i) a flexible network-on-chip (NoC) supporting multi-dataflow and sparsity on precision-scalable MAC array, and ii) efficient data storage using an optimal sparsity format based on the sparsity ratio and precision modes.
To evaluate the effectiveness of FlexNeRFer, we performed a layout implementation using 28nm CMOS technology.
Our evaluation shows that FlexNeRFer achieves 8.2$\sim$243.3$\times$ speedup and 24.1$\sim$520.3$\times$ improvement in energy efficiency over a GPU (i.e., NVIDIA RTX 2080 Ti), while demonstrating 4.2$\sim$86.9$\times$ speedup and 2.3$\sim$47.5$\times$ improvement in energy efficiency compared to a state-of-the-art NeRF accelerator (i.e., NeuRex).
\end{abstract}



\maketitle

\section{Introduction}\label{sec:intro}

\begin{table}[t]
\centering
\caption{Design specifications of modern GPU devices used in on-device rendering.} 
\label{tab:gpu_spec}\vspace{-2mm}
\scalebox{0.79}{
\begin{tabular}{c cc cc}
\hline
                                                                       & \multicolumn{2}{c}{\textbf{Desktop GPU}}                                                                                       & \multicolumn{2}{c}{\textbf{Edge GPU}}                                                                                             \\ \hline\hline
\textbf{GPU Model}                                                        & \textbf{RTX 2080 Ti}                                   & \textbf{RTX 4090}                                    & \textbf{Jetson Nano}                                  & \textbf{Xaiver NX}                                    \\ \hline
\textbf{Process Node {[}nm{]}}                                      & 12                                                  & 5                                                    & 20                                                    & 12                                                    \\ \hline
\textbf{Area {[}mm$^2${]}}                                                & 754                                                 & 609                                                  & 118                                                   & 350                                                   \\ \hline
\textbf{Frequency {[}GHz{]}}                                               & 1.4                                                 & 2.3$\sim$2.6                                        & 0.9                                                   & 1.1                                                   \\ \hline
\textbf{Typical Power {[}W{]}}                                                 & 250                                                 & 350                                                  & 10                                                    & 20                                                    \\ \hline
\textbf{\begin{tabular}[c]{@{}c@{}}DRAM Bandwidth \\ {[}GB/s{]}\end{tabular}} & \begin{tabular}[c]{@{}c@{}}GDDR6\\ 616\end{tabular} & \begin{tabular}[c]{@{}c@{}}GDDR6\\ 1150\end{tabular} & \begin{tabular}[c]{@{}c@{}}LPDDR4\\ 25.6\end{tabular} & \begin{tabular}[c]{@{}c@{}}LPDDR4\\ 59.7\end{tabular} \\ \hline
\end{tabular}}\vspace{-2mm}
\end{table}

Neural Radiance Fields (NeRF) have significantly advanced the field of 3D scene reconstruction by enabling smooth and high-resolution view synthesis through neural networks~\cite{grid_representation, deep_sdf, local_deep_implict, occupancy_network}. 
Unlike traditional methods that reconstruct scenes using explicit representations such as meshes or voxel grids, NeRF models employ multi-layer perceptrons (MLPs) which map spatial information (e.g., xyz coordinates) to radiance values, reproducing complex geometric structures and lighting conditions with photorealistic reconstruction quality~\cite{vanilar_nerf}. 
This capability has positioned NeRF as a cornerstone technology across various applications, including 3D asset creation for the metaverse~\cite{3d_asset}, VR painting~\cite{vr_painting}, video games~\cite{luma_video}, and autonomous vehicle perception~\cite{nerf_car}.

As NeRF adoption grows, various model variants have been developed to meet diverse performance and quality requirements. 
For instance, in applications such as gaming, where real-time rendering is critical, NeRF models optimized for computational speed have been developed using approaches such as simplified MLP structures~\cite{instant_ngp, kilonerf, video2game, bitrate_nf}, data quantization~\cite{ptq_nerf, tinynerf, how_far_quantization, bitrate_nf}, or sparsity techniques (e.g., pruning and sparse representation)~\cite{tinynerf, hollonerf, plenoxel, masked_wavelet, meta_learning_pruning}. 
In contrast, in applications where high-resolution detail is crucial, such as architectural visualization and medical education, NeRF models utilizing high-precision multiscale data representations are used~\cite{mip_nerf, TensoRF, Plenoctree}.

\begin{figure}[t]
    \centering
    \includegraphics[scale=0.77]{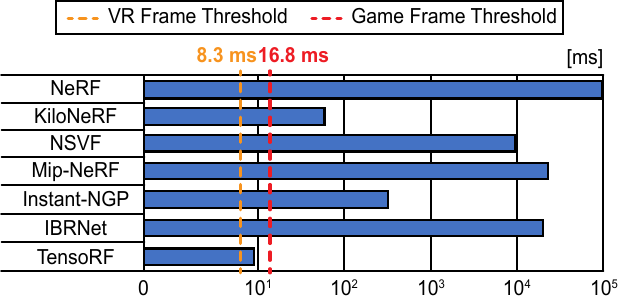}
    \caption{Rendering latency of seven representative NeRF models, i.e., NeRF~\cite{vanilar_nerf}, KiloNeRF~\cite{kilonerf}, NSVF~\cite{NSVF}, Mip-NeRF~\cite{mip_nerf}, Instant-NGP~\cite{instant_ngp}, IBRNet~\cite{ibrnet}, and TensoRF~\cite{TensoRF}, on the NVIDIA RTX 2080 Ti using the Synthetic-NeRF dataset~\cite{synthetic_nerf}.}
\label{fig:moti_rendering_time}\vspace{-2mm}
\end{figure}

However, devices typically used to run various NeRF-based applications, i.e., GPUs, have limitations due to their large chip area and high power consumption, making them unsuitable for portable devices.
As shown in Table~\ref{tab:gpu_spec}, modern desktop-scale and edge GPUs from NVIDIA do not meet the stringent area (less than 100~mm\textsuperscript{2}) and power constraints (under 10~W) necessary for devices like AR glasses \cite{ar_glass, google_glass, hololense}. 
Moreover, GPUs often fail to deliver the necessary performance for real-time applications. 
According to the studies in~\cite{game_constraint} and~\cite{post_vr}, VR and video games are required to maintain a frame time latency below 16.8ms and 8.3ms, respectively, to prevent issues such as cybersickness, eye strain, and reduced immersion.
Fig.~\ref{fig:moti_rendering_time} shows that the rendering times of seven representative NeRF models~\cite{vanilar_nerf, kilonerf, NSVF, mip_nerf, instant_ngp, ibrnet, TensoRF} on an NVIDIA RTX 2080 Ti GPU exceed the frame time thresholds.

To address these limitations, several NeRF accelerators have been proposed~\cite{icarus, metavrain, instant_3d, neu_gpu, neurex, instant_nerf_dac, ngpo, hi_nerf, gen_nerf, rf_nerf_acc}.
However, existing accelerator solutions are primarily optimized for specific NeRF models, limiting their adaptability and efficiency across different models.
One possible approach to overcoming this limitation is equipping devices with specialized accelerators for each NeRF model.
Yet, this remains impractical, especially on portable devices, due to constraints in production cost, size, and energy efficiency~\cite{hetero_acc_limit}.
Furthermore, the rapid evolution of NeRF models necessitates a hardware accelerator capable of adapting to new models and algorithms without redesign or redeployment. 
It creates a pressing need for a flexible hardware accelerator capable of efficiently supporting the diverse requirements of NeRF models in real-world deployments.

Designing a flexible NeRF accelerator introduces several technical challenges. 
First, \textit{diverse computational workloads} across different NeRF models, which stem from variations in neural network architectures such as MLP (for coordinate-based regression)~\cite{vanilar_nerf, vanilar_nerf, kilonerf, video2game, bitrate_nf}, CNN (for image-based feature extraction)~\cite{gen_nerf_transformer, pixel_nerf, nerf_rpn, vit_nerf, ha_nerf}, and Transformer (for attention-based scene representation)~\cite{gen_nerf_transformer, contra_nerf, able_nerf, vit_nerf, nerf_rpn}, as well as different data encodings (e.g., positional encoding~\cite{vanilar_nerf, kilonerf, NSVF, mip_nerf} and hash encoding~\cite{instant_ngp, gen_nerf_transformer, intrinsic_ngp, camera_pose_hash}), require the accelerator to process these heterogeneous workloads efficiently.
This demands adaptable computation units that can adjust to the specific needs of each model without sacrificing performance. 
Second, many NeRF models employ \textit{sparsity techniques} such as pruning~\cite{gen_nerf, pru_nerf} and sparse voxel filtering~\cite{NSVF, Plenoctree, hollonerf, tinynerf} to reduce computational overhead. 
Efficiently leveraging sparsity necessitates hardware support for zero-skipping computation and compressed representation of sparse data.
Third, different models and applications require \textit{varying levels of numerical precision} to trade off computational accuracy and efficiency. 
Supporting multiple precision modes (e.g., 4-bit~\cite{tinynerf, bitrate_nf}, 8-bit~\cite{compgs, compress_1mb}, and 16-bit~\cite{Plenoctree, booth_nerf}) within a single hardware framework is essential for optimizing performance and energy consumption.

Although these challenges seem to be conquered by the prior work on deep learning accelerators, they lack providing a holistic solution that covers all these challenges in a single architecture.
To do so, we first analyze the computational steps that cause performance bottlenecks in representative NeRF algorithms, when executed on the off-the-shelf GPU (i.e., NVIDIA RTX 2080 Ti~\cite{rtx_2080_ti}) [Section~\ref{sec:performance_profiling}].
Subsequently, to mitigate these performance bottlenecks, we explore the use of commercial accelerators (e.g., Google TPU~\cite{tpu_v4} and NVIDIA NVDLA~\cite{nvdira}) [Section~\ref{sec:inefficiency_dense_acc}]. 
However, our investigation shows that these accelerators still suffer from suboptimal performance, leading us to derive key design requirements to overcome these constraints [Section~\ref{sec:sec_need_dataflow}].
Furthermore, for the first time, we examine primary design challenges associated with handling sparsity in a precision-scalable 2D MAC array designed to support quantization [Section~\ref{sec:need_adaptive_sparsity}].

\begin{figure*}[t]
    \centering
    \includegraphics[scale=0.56]{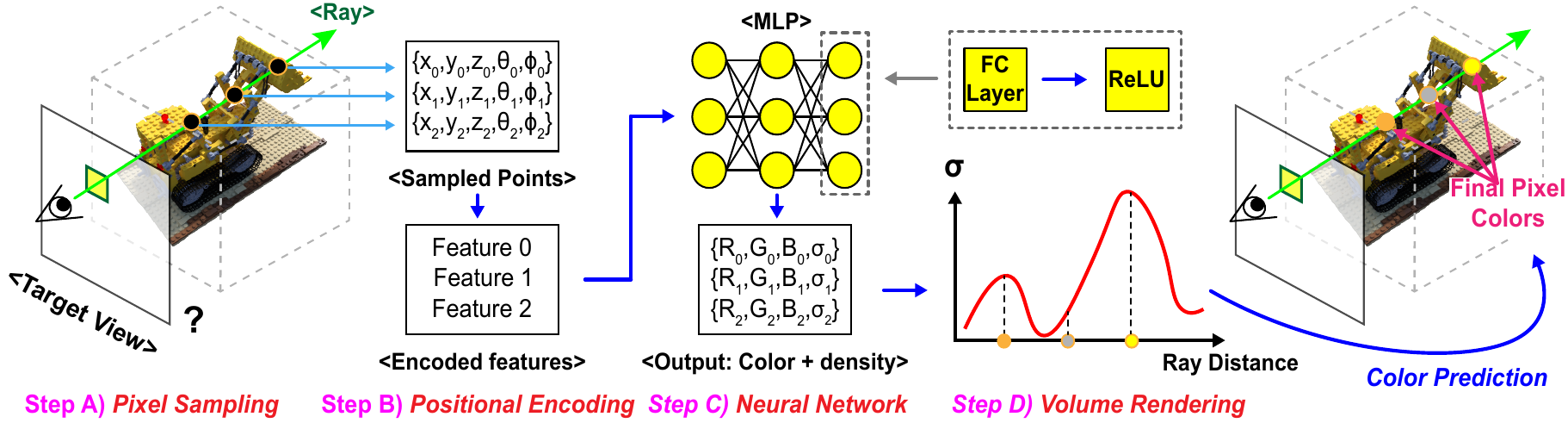}\vspace{-2mm}
    \caption{Visual guide to the execution pipeline of NeRF~\cite{vanilar_nerf}.}
    \label{fig:background_nerf_pipeline}\vspace{-2mm}
\end{figure*}

Based on these analyses, this paper proposes \textit{\underline{FlexNeRFer}}, a flexible on-device NeRF accelerator designed to efficiently run various NeRF models with low power-performance-area (PPA) costs.
The FlexNeRFer includes an encoding unit and a GEMM/GEMV acceleration unit, both of which accelerate neural feature encoding and GEMM/GEMV computations, respectively, which are key performance bottlenecks in NeRF models [Section~\ref{sec:flexnerfer_archi}].
In particular, the GEMM/GEMV acceleration unit features a hierarchical and flexible distribution network that supports various dataflows (unicast, multicast, and broadcast) to efficiently map sparse data onto the precision-scalable MAC array along the 2D direction in a dense manner [Section~\ref{sec:distribution_noc}].
In addition, it integrates a reduction tree that dynamically aggregates data within the 2D structure to optimize compute efficiency [Section~\ref{sec:reduction_noc}].
Furthermore, based on the observation that the data compression format minimizing memory footprint varies with precision mode and sparsity ratio, FlexNeRFer supports sparsity-aware data compression, which dynamically identifies the sparsity ratio of input data in real time while pre-analyzing that of weight data, adopting the optimal data format accordingly  [Section~\ref{sec:data_compression}].
With these architectural features, FlexNeRFer enables efficient scene generation for various NeRF models on-device while maintaining low hardware cost [Section~\ref{sec:evaluation}].

Our key contributions are as follows:\vspace{-1mm}
\begin{enumerate}
    \item \textbf{Design Requirement Analysis:} To enable broad acceleration of NeRF models, we first identify common performance bottlenecks across various NeRF algorithms and establish design requirements to address these challenges. In addition, we explore strategies for efficiently supporting sparsity in a bit-flexible MAC array.
    \item \textbf{Flexible Network-on-Chip (NoC):} We introduce a flexible and hierarchical NoC architecture that enhances the performance of common bottleneck operations in NeRF models on a precision-scalable MAC array.
    \item \textbf{Online Sparsity-Aware Data Compression:} We propose a framework that determines the sparsity ratio of input data in real-time on a per-tile basis while pre-analyzing weight data. The tensors are then encoded into an optimal sparsity format based on precision and sparsity ratio, minimizing memory usage. 
    \item \textbf{Thorough Architectural Evaluation:} We evaluate the proposed accelerator by compariring it against state-of-the-art accelerators and a consumer-grade GPU across seven NeRF models to assess its effectiveness.
\end{enumerate}


\section{Background}\label{sec:background}

\subsection{Preliminary of NeRF}\label{sec:preliminary}

\subsubsection{Neural Radiance Fields (NeRF)}\label{sec:nerf_concept}
\textbf{\newline}The concept of the neural radiance fields (NeRF) was introduced by Mildenhall et al. for 3D view reconstruction~\cite{vanilar_nerf}. 
This work presented a neural network-based rendering method that predicts a set of 3D points for novel views not included in the training dataset. 
Due to its ability to achieve photo-realistic reconstruction quality in complex scenes, it has driven the development of advanced NeRF models, which have now become a compelling strategy for various vision tasks~\cite{grid_representation, deep_sdf, local_deep_implict, occupancy_network, instant_3d}. 
These advanced models build upon the fundamental concept of the vanilla NeRF. 
Thus, we outline how rendering is performed in the vanilla NeRF.

Fig.~\ref{fig:background_nerf_pipeline} depicts the rendering pipeline of the vanilla NeRF. 
It involves four main steps: A) pixel sampling, B) positional encoding, C) neural network processing, and D) volume rendering.
\textbf{[Step A: Pixel sampling]} The first step, pixel sampling, entails generating rays for each pixel of the target object and sampling points along those rays. 
The ray provides a 5D representation for every sampled point, consisting of spatial coordinates (x, y, z) and azimuthal and polar viewing angles ($\phi$, $\theta$), using ray marching.
\textbf{[Step B: Positional encoding]} Next, the 5D feature is encoded into a high-dimensional vector ($\gamma(v)$) using sinusoidal positional encoding~\cite{positional_encoding, fourier_encoding}. 
This encoding enables the neural network to capture high-frequency signals more effectively, which is computed as follows:\vspace{-0.5mm}
\begin{equation}\label{eq:postional_encoding}
\gamma(v) = \{\sin(2^0 \pi v), \cos(2^0 \pi v), \ldots, \cos(2^{N-1} \pi v)\},
\end{equation}
where $v$ represents one of the 5D features (i.e., x, y, z, $\phi$, $\theta$), and $N$ is a hyper-parameter that determines the encoding dimensionality.
\textbf{[Step C: Neural network]} Afterward, the neural network computes the color (c: R, G, B) and density ($\sigma$) for each sampled point.
\textbf{[Step D: Volume Rendering]} Finally, the color of each pixel ($C(\mathbf{r})$) is obtained by integrating the color and density values along the ray for all sampled points, weighted by their transmittance ($T_i$):
\begin{equation}\label{eq:integral_eq}
C(\mathbf{r}) = \int_{t_1}^{t_2} T(t) \cdot \sigma(t) \cdot c(t) \, dt,
\end{equation}
where $T(t) = \exp \left( - \int_{t_n}^{t} \sigma(u) \, du \right)$ denotes the accumulated transmittance, which represents the probability that the ray travels from $t_n$ to $t$ without being intercepted. 
In practice, Eq.~(\ref{eq:integral_eq}) is approximated using numerical quadrature by sampling points:
\begin{equation}
\hat{C}(\mathbf{r}) = \sum_{i=1}^{N} T_i (1 - \exp\left(-\sigma_i \delta_i\right))  c_i,
\vspace{-1mm}\end{equation}
where $T_i = \exp\left(- \sum_{j=1}^{i-1} \sigma_j \delta_j\right)$, and $\delta_i$ represents the distance from sample $i$ to sample $i + 1$.

\subsubsection{NeRF Acceleration vs. DNN (LLM) Acceleration}\label{sec:nerf_vs_ai}

\textbf{\newline}
As the demand for Transformer-based LLMs continues to grow across various real-world applications, research efforts in both academia and industry have been actively directed toward improving inference speed~\cite{early_llm_acc, dota_early_llm, llmcompass, samsung_llm, npu_pim_llm, splitwise_llm, softmax_moe, parameter_moe, flashdecoding, pipe_moe}.
In early LLMs, large-scale matrix operations in self-attention and feedforward networks constituted the majority of the computational workload, making their efficient acceleration a primary research objective~\cite{early_llm_acc, dota_early_llm}.
One emerging concern in LLM acceleration is the Softmax function, which is used in every attention block to compute attention scores.
In addition, recent LLMs have adopted the Mixture-of-Experts (MoE) technique, which selectively activates a subset of expert networks based on the input~\cite{original_moe, routing_moe} to reduce computational costs and enhance inference speed.
In MoE-based LLMs, Softmax is used not only for computing attention scores but also for expert selection, increasing its computational overhead~\cite{softmax_moe, parameter_moe}.
As a result, recent DNN acceleration research has focused on optimizing the execution of large-scale matrix operations (GEMM/GEMV) alongside Softmax to minimize computational costs and improve inference speed~\cite{flashdecoding, pipe_moe}.

NeRF utilizes neural networks, typically using MLPs, to predict the color (R, G, B) and density ($\sigma$) of 3D coordinates sampled along rays~\cite{nerf_survey}. 
While NeRF shares similarities with DNNs in that it employs neural networks for specific tasks, it differs by incorporating additional operations such as neural feature encoding and volume rendering. 
In this paper, we analyze critical common bottlenecks across various NeRF models, identifying \textit{GEMM/GEMV operations} and \textit{encoding processes} as key performance limitations [Section~\ref{sec:performance_profiling}]. 
To address these challenges, we introduce FlexNeRFer, which features dedicated modules specifically designed to optimize these operations [Section~\ref{sec:architecture_feature} and~\ref{sec:flexnerfer_archi}].
Furthermore, the varying sparsity of tensors in NeRF models makes our FlexNeRFer well-suitable for neural rendering, benefiting from its online sparsity-aware data compression.
Note that GEMM/GEMV acceleration techniques proposed in FlexNeRFer are not limited to NeRF workloads but are also applicable to general DNN/LLM accelerators.
Thus, the novel design components of FlexNeRFer extend beyond NeRF workloads, offering broader applicability in general deep learning acceleration.


\begin{table}[t]
\centering
\caption{
Comparison between FlexNeRFer and related works, exploring flexible network-on-chip (NoC) in terms of versatile dataflow, multi-sparsity format, and bit-level flexibility. In the table, 'U', 'M', and 'B' represent unicast, multicast, and broadcast, respectively, while 'IP', 'OP', and 'RP' denote inner product, outer product, and row-wise product, respectively.
} 

\label{tab:comparison_noc}
\scalebox{0.73}{
\begin{tabular}{|c|c|c|c|}
\hline
\textbf{Related Work} & \textbf{\begin{tabular}[c]{@{}c@{}}Dataflow \\ Flexibility\\ (Dataflow Modes)\end{tabular}} & \textbf{\begin{tabular}[c]{@{}c@{}}Multi-Sparsity\\ Format\\ (Supported Format)\end{tabular}} & \textbf{\begin{tabular}[c]{@{}c@{}}Bit-level \\ Flexibility\\ (Data Bit-Width)\end{tabular}} \\ \hline\hline
\textbf{Microswitch~\cite{microswitch}}  & \begin{tabular}[c]{@{}c@{}}\cmark \\ ( U, M, B )\end{tabular}                           & \begin{tabular}[c]{@{}c@{}}\xmark \\ ( N/A )\end{tabular}                                            & \begin{tabular}[c]{@{}c@{}}\xmark \\ ( - )\end{tabular}                                 \\ \hline
\textbf{Eyeriss v2~\cite{eyeriss_v2}}   & \begin{tabular}[c]{@{}c@{}}\cmark \\ ( U, M, B )\end{tabular}                           & \begin{tabular}[c]{@{}c@{}}\xmark \\ ( N/A )\end{tabular}                                            & \begin{tabular}[c]{@{}c@{}}\xmark \\ ( 8 )\end{tabular}                                             \\ \hline
\textbf{SIGMA~\cite{sigma}}        & \begin{tabular}[c]{@{}c@{}}\cmark \\ ( U, M, B )\end{tabular}                           & \begin{tabular}[c]{@{}c@{}}\xmark \\ ( Bitmap )\end{tabular}                                         & \begin{tabular}[c]{@{}c@{}}\xmark \\ ( 16 )\end{tabular}                                            \\ \hline
\textbf{Flexagon~\cite{flexagon}}     & \begin{tabular}[c]{@{}c@{}}\cmark \\ ( IP, OP, RP )\end{tabular}                 & \begin{tabular}[c]{@{}c@{}}\xmark \\ ( CSC / CSR )\end{tabular}                                       & \begin{tabular}[c]{@{}c@{}}\xmark \\ ( - )\end{tabular}                                 \\ \hline
\textbf{Trapezoid~\cite{trapezoid}}      & \begin{tabular}[c]{@{}c@{}}\cmark \\ ( IP, RP )\end{tabular}                            & \begin{tabular}[c]{@{}c@{}}\xmark \\ ( CSC / CSR )\end{tabular}                                           & \begin{tabular}[c]{@{}c@{}}\xmark \\ ( 32 )\end{tabular}                                             \\ \hline
\textbf{FEATHER~\cite{feather}}      & \begin{tabular}[c]{@{}c@{}}\cmark \\ ( U, M,  B )\end{tabular}                           & \begin{tabular}[c]{@{}c@{}}\xmark \\ ( N/A )\end{tabular}                                            & \begin{tabular}[c]{@{}c@{}}\xmark \\ ( 8 )\end{tabular}                                             \\ \hline
\textbf{\begin{tabular}[c]{@{}c@{}}FlexNeRFer \\ (Ours)\end{tabular}    }   & \begin{tabular}[c]{@{}c@{}}\cmark \\ ( U, M, B )\end{tabular}                           & \begin{tabular}[c]{@{}c@{}}\cmark \\ ( CSC / CSR \\( COO, and Bitmap )\end{tabular}                & \begin{tabular}[c]{@{}c@{}}\cmark \\ ( 4, 8, 16 )\end{tabular}                                       \\ \hline
\end{tabular}
}
\end{table}

\subsection{Related Work}\label{sec:related_work}
\subsubsection{NeRF Accelerator}\label{sec:related_work_nerf_acc}
\textbf{\newline}To overcome the inefficiencies of rendering using NeRF models on GPU devices, many research efforts have been made recently to accelerate NeRF computation in an energy-efficient manner.
For instance, ICARUS~\cite{icarus} and MetaVRain~\cite{metavrain} proposed accelerator designs consisting of custom hardware modules optimized for each computational process of vanilla NeRF~\cite{vanilar_nerf}.
Additionally, several studies such as~\cite{instant_3d, neu_gpu, neurex, instant_nerf_dac, ngpo, hi_nerf, srender} and~\cite{fusion_3d} have presented hardware architectures that accelerate the bottleneck processes of Instant-NGP~\cite{instant_ngp}, specifically parametric encoding (i.e., hash encoding) and the feature querying process along rays using an MLP, aiming to achieve real-time rendering on mobile and edge devices.
Moreover, two pioneering works,~\cite{gen_nerf} and~\cite{rf_nerf_acc}, introduced algorithm-hardware co-design approaches tailored for IBRNet~\cite{ibrnet} and TensoRF~\cite{TensoRF}, respectively, focusing on point sampling and memory efficiency. 
However, these prior studies focused on specific NeRF models, indicating that their remarkable acceleration performance may not generalize to other models (\textbf{\textit{Limitation of related works}}). 
In this paper, we investigate the processes that lead to bottlenecks in various NeRF models and propose a hardware architecture that universally achieves high performance with low hardware cost across diverse NeRF models.

\subsubsection{Flexible Network-on-Chip (NoC)}\label{sec:related_work_noc}
\textbf{\newline}Various studies have been conducted on flexible NoC to improve performance by maximizing hardware resource utilization. 
For instance, Microswitch~\cite{microswitch} and Eyeriss v2~\cite{eyeriss_v2} identified the necessity of supporting diverse dataflows (e.g., unicast, multicast, and broadcast) when performing convolution operations on spatial MAC units and proposed flexible NoC structures to meet this need. 
SIGMA employed a Benes network and a forwarding adder network to ensure high MAC utilization in weight stationary systolic array (SA) for sparse and irregular GEMM operations~\cite{sigma}.
Flexagon~\cite{flexagon} and Trapezoid~\cite{trapezoid}, observing that optimal dataflow (e.g., inner product, output product, row-wise product) varies with the sparsity ratio of data, proposed a reconfigurable interconnect to support these dataflows and enhance performance.
Finally, FEATHER introduced an interconnect architecture that supports data reordering and a range of dataflows, including unicast, multicast, and broadcast, ensuring efficient dataflow switching~\cite{feather}.
However, none of the previous studies have considered multiple data bit-widths (\textbf{\textit{Unexplored issue}}).
Based on the observation that the optimal sparsity format for achieving a low memory footprint varies with the bit-width of data [Section~\ref{sec:need_adaptive_sparsity}], we propose a flexible NoC structure that can not only support various dataflows and irregular GEMM/GEMV operations, but also handle multiple sparsity formats.
Table~\ref{tab:comparison_noc} summarizes a comparison with related works\footnote{
Since CSC and CSR share the same compression mechanism~\cite{flexagon, linear_sparse_system, gamma}, differing only in whether the data is stored row-wise or column-wise, they are grouped into a single category.}.

\begin{figure}[t]
    \centering
    \includegraphics[scale=0.77]{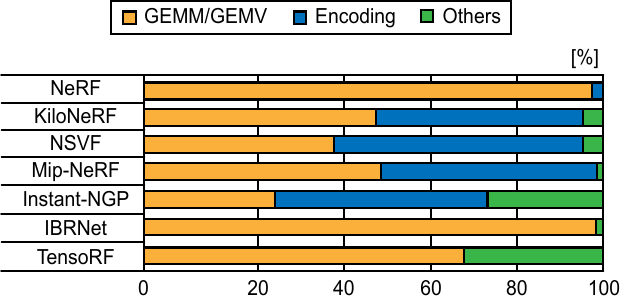}
    \caption{Runtime breakdown of seven modern NeRF models, including NeRF~\cite{vanilar_nerf}, KiloNeRF~\cite{kilonerf}, NSVF~\cite{NSVF}, Mip-NeRF~\cite{mip_nerf}, Instant-NGP~\cite{instant_ngp}, IBRNet~\cite{ibrnet}, and TensoRF~\cite{TensoRF}, evaluated on an NVIDIA RTX 2080 Ti using the Synthetic-NeRF dataset~\cite{synthetic_nerf}. In the figure, when encoding is performed using GEMM/GEMV operations, its execution time is included within the GEMM/GEMV runtime.}
    \label{fig:runtime_break}\vspace{-2mm}
\end{figure}

\begin{figure*}[t]
    \centering
    \includegraphics[scale=0.63]{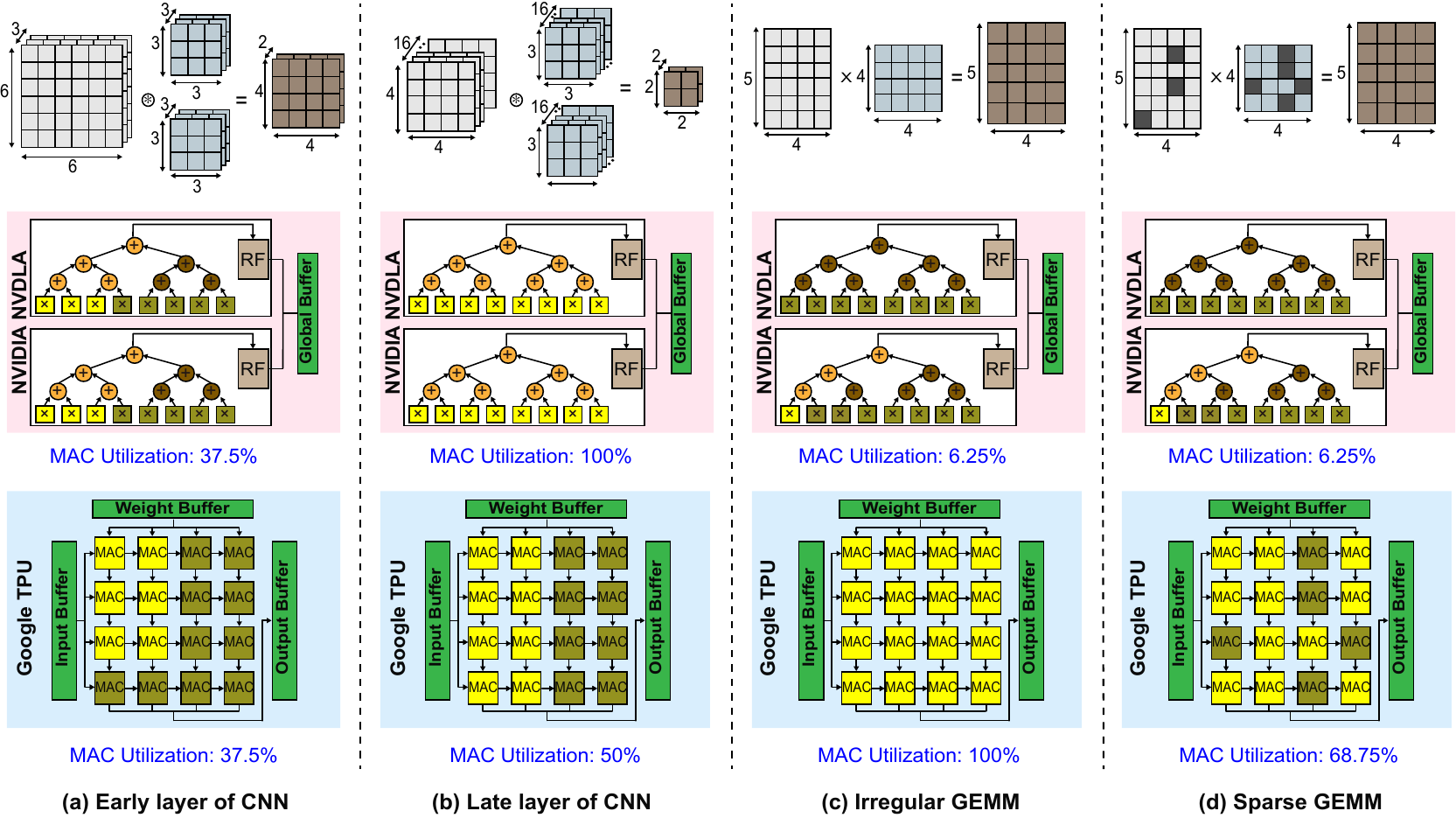}\vspace{0mm}
    \caption{MAC utilization of two commercial accelerators, i.e., NVIDIA NVDLA~\cite{nvdira} and Google TPU~\cite{tpu_v4}, across various scenarios. (a) and (b) illustrate MAC utilization when data from the early and late layers of CNNs are mapped to two accelerators, respectively. (c) and (d) present MAC utilization during irregular dense and sparse GEMM operations, which form the backbone of computations in MLPs and Transformers.}
    \label{fig:mac_uti}
\end{figure*}

\section{Design Requirements for NeRF Acceleration}\label{sec:motivation}
Neural rendering with NeRF demands substantial computational intensity. 
For example, the original NeRF model requires 18,000$\times$ and 23,700$\times$ more operations than ResNet-50~\cite{resnet} and Cycle-GAN~\cite{cyclegan}, respectively~\cite{metavrain_journal}. 
This immense computational load poses significant challenges for real-time processing and low-power implementation on on-device systems.
In response, numerous hardware accelerators specifically optimized for NeRF operations have been proposed in recent years~\cite{icarus, metavrain, instant_3d, neu_gpu, neurex, instant_nerf_dac, ngpo, hi_nerf, gen_nerf, rf_nerf_acc, srender, fusion_3d}. 
However, these accelerator designs are optimized to individual NeRF algorithms, thereby limiting their efficiency across different NeRF models.
To address these limitations, this section profiles the performance of multiple NeRF models to identify common performance bottlenecks. 
From this insight, we derive key hardware requirements necessary to alleviate these constraints. 
Additionally, we observe that the optimal compressed format for minimizing memory usage depends on both the precision mode and sparsity ratio; accordingly, we specify detailed hardware requirements to effectively support dynamic sparsity formats.

\subsection{Performance Profiling}\label{sec:performance_profiling}
To identify common bottlenecks in NeRF models, we analyzed the runtime breakdown using a NVIDIA RTX 2080 Ti GPU, which is typically employed for 3D view reconstruction, across seven representative NeRF models from various applications: NeRF~\cite{vanilar_nerf}, KiloNeRF~\cite{kilonerf}, NSVF~\cite{NSVF}, Mip-NeRF~\cite{mip_nerf}, Instant-NGP~\cite{instant_ngp}, IBRNet~\cite{ibrnet}, and TensoRF~\cite{TensoRF}.
Fig.~\ref{fig:runtime_break} presents the runtime breakdown divided into the respective proportions of GEMM/GEMV operations, encoding, and other computations.
As shown, GEMM/GEMV operations account for a significant portion of the total execution time across all NeRF models. 
Furthermore, in certain NeRF models (e.g., KiloNeRF~\cite{kilonerf}, NSVF~\cite{NSVF}, Mip-NeRF~\cite{mip_nerf}, and Instant-NGP~\cite{instant_ngp}), considerable latency is consumed in the encoding process, where spatial coordinates are transformed into higher-dimensional representations.
Therefore, to enable universally fast rendering using NeRF models, NeRF accelerators must efficiently optimize and accelerate GEMM/GEMV operations. 
Moreover, effective support for encoding is crucial to ensure consistently high performance across a range of NeRF models \textbf{\textit{(Takeway 1)}}.
Based on this observation, the following subsection provides an in-depth examination of the design requirements for accelerating GEMM/GEMV operations. 
In addition, a brief discussion of hardware modules for accelerating encoding process is provided in [Section~\ref{sec:nerf_encoding_unit}].

\begin{figure}[t]
    \centering
    \includegraphics[scale=0.76]{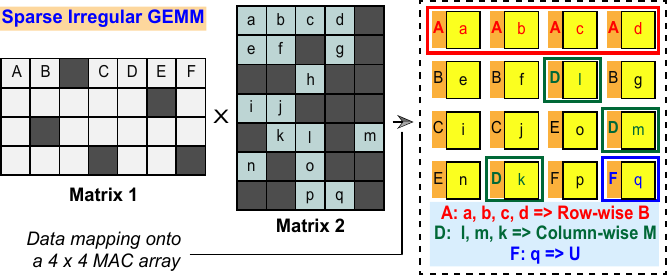}
    \caption{Data mapping of an irregular sparse GEMM operation onto a 4×4 MAC array in a dense manner. `B', `M', and `U' denote broadcast, multicast, and unicast, respectively.}
    \label{fig:moti_dataflow}
\vspace{-2mm}
\end{figure}

\subsection{Acceleration of GEMM/GEMV Operations}\label{sec:gem_gemv_acceleration}

\subsubsection{Inefficiency of Dense GEMM/GEMV Accelerators}\label{sec:inefficiency_dense_acc}
\textbf{\newline}NeRF models employ various neural architectures to generate high-quality novel views without the need to construct complex geometric structures.
For instance, NeRF eliminates the need for explicit 3D representations such as meshes and voxels by leveraging an implicit neural representation, where an MLP maps spatial coordinates and viewing directions to color and density values~\cite{vanilar_nerf, vanilar_nerf, kilonerf, video2game, bitrate_nf}.
Some NeRF models utilize CNNs to extract and aggregate features from multiple camera views and different rays, improving the accuracy of reconstructed 3D scenes~\cite{gen_nerf_transformer, pixel_nerf, nerf_rpn, vit_nerf, ha_nerf}.
Additionally, certain NeRF variants incorporate a ray transformer to assign weights to extracted features based on their relevance to the target ray, enhancing the realism of rendered scenes~\cite{gen_nerf_transformer, contra_nerf, able_nerf, vit_nerf, nerf_rpn}.
While these neural architectures enable high-quality rendering, they also introduce significant computational overhead, particularly in GEMM/GEMV operations, which are a common bottleneck across various NeRF models, as analyzed in [Section~\ref{sec:performance_profiling}].
To address this challenge, prior research has explored sparsity techniques as a means to reduce computational cost.
For instance, the works~\cite{NSVF, Plenoctree, hollonerf} and ~\cite{tinynerf} exploit sparse voxel grids by filtering samples from unbounded regions or occluded areas behind target objects.
Prior studies by~\cite{gen_nerf} and~\cite{pru_nerf} apply structured pruning to reduce the number of GEMM/GEMV operations.
Therefore, for NeRF accelerators to speed up GEMM/GEMV operations, the modules designed to process these computations must efficiently support various types of neural networks and sparsity \textbf{\textit{(Takeaway 2)}}.

While commercial GEMM/GEMV accelerators can be utilized to accelerate these operations, their specialized designs for specific dense neural network workloads limit high performance across diverse neural network computations and sparse operations.
Fig.~\ref{fig:mac_uti} illustrates how various neural network operations and sparsity affect MAC utilization on two commercial accelerators: NVIDIA NVDLA~\cite{nvdira} and Google TPU~\cite{tpu_v4}.
Specifically, Fig.~\ref{fig:mac_uti}-(a) and (b) present the MAC utilization when mapping data from the early and late layers of a CNN, respectively, and Fig.~\ref{fig:mac_uti}-(c) and (d) depict the MAC utilization for irregular dense and irregular sparse GEMM operations.
In the early layers of a CNN, both accelerators show low utilization due to the shallow channel depth (Fig.~\ref{fig:mac_uti}-(a)). 
In the late layers, NVDLA achieves 100\% MAC utilization, whereas Google TPU continues to have low MAC utilization (Fig.~\ref{fig:mac_uti}-(b)). 
For irregular dense GEMM operations, Google TPU achieves high utilization for the given matrix dimensions, while NVDLA demonstrates significantly lower utilization (Fig.~\ref{fig:mac_uti}-(c)). 
MAC utilization decreases further in Google TPU due to to data sparsity in irregular sparse GEMM operations (Fig.~\ref{fig:mac_uti}-(d)).
The low MAC utilization leads to a reduction in throughput, resulting in performance degradation during GEMM/GEMV operations. 
Thus, an ideal NeRF accelerator that delivers improved performance across various NeRF models must maintain high hardware resource utilization for various neural network operations and sparse data \textbf{\textit{(Design requirement 1)}}. 

\vspace{1mm}
\subsubsection{Need for Multi-dataflow Support}\label{sec:sec_need_dataflow}
\textbf{\newline}In this section, we examine the design requirement for an interconnect that enables dense data mapping onto MAC units, ensuring consistently high MAC utilization across different neural network operations and varying levels of sparsity.
Fig.~\ref{fig:moti_dataflow} illustrates how data from an irregular sparse GEMM operation is densely mapped onto a 4 $\times$ 4 MAC array. 
In the first row of the MAC array, the red box indicates that operand ‘A’ from matrix 1 is broadcast row-wise.
The blue boxes in the figure represent MAC units where mapped data from matrix 2 shares the same element ‘B’ from matrix 1 along the column direction.
Finally, the green box in the last row represents a MAC unit that performs a MAC operation on 'F' from matrix 1 and 'q' from matrix 2.
As shown, achieving a dense mapping of GEMM/GEMV elements onto the MAC array requires the support of broadcast, multicast, and unicast dataflows in both row and column directions \textbf{\textit{(Takeway 3; Design requirement 2)}}. 

\begin{figure}[t]
    \centering
    \includegraphics[scale=0.67]{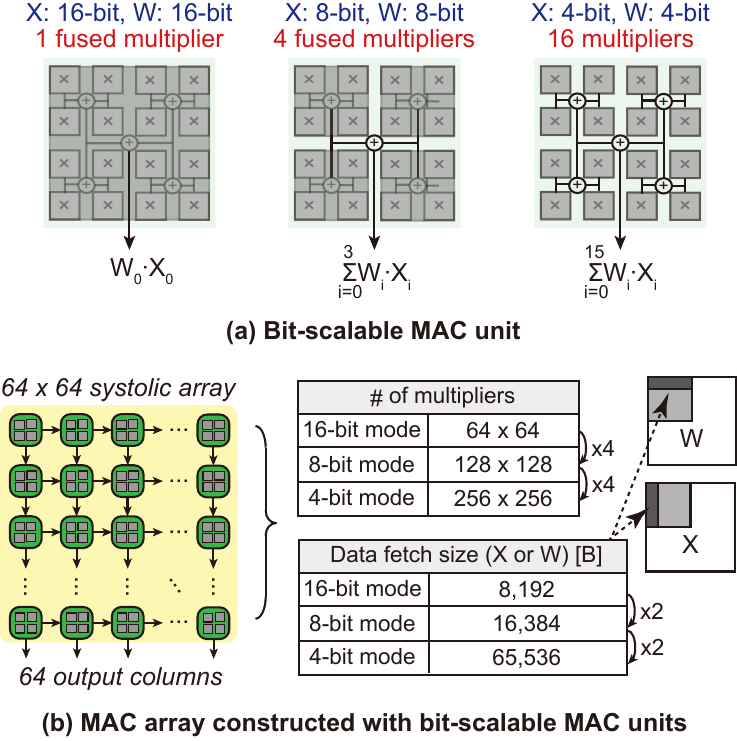}
    \caption{(a) Operations of a bit-scalable MAC unit in 4-bit, 8-bit, and 16-bit modes. (b) A MAC array consisting of bit-scalable MAC units, showing multiplier counts and tile sizes for different bit-widths.}
    \label{fig:fu}
\end{figure}

\begin{figure*}[t]
    \centering
    \includegraphics[scale=0.62]{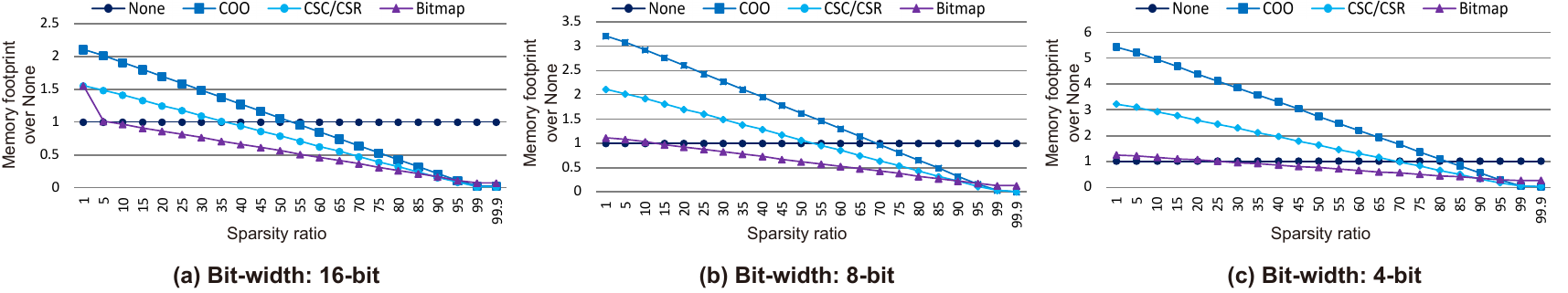}\vspace{-2mm}
    \caption{Memory footprints of various compression formats (COO, CSC/CSR, Bitmap) normalized to `None' (i.e., uncompressed) across different sparsity ratios for 16-bit  (matrix size: 64 $\times$ 64), 8-bit (matrix size: 128 $\times$ 128), and 4-bit (matrix size: 256 $\times$ 256) precision modes.}
    \label{fig:sparsity_ratio}
\end{figure*}

\vspace{1mm}
\subsubsection{Necessity of Adaptive Sparsity Format Support}\label{sec:need_adaptive_sparsity}
\textbf{\newline}Building on recent research efforts in data quantization and pruning for NeRFs, FlexNeRFer supports multiple bit-widths and sparsity formats.
According to the study in~\cite{survey19}, which provides a comparative analysis of MAC units supporting various precision levels, the bit-scalable MAC unit introduced in~\cite{bitfusion} is the most energy-efficient when designing MAC arrays.
Based on this finding, FlexNeRFer adopts the MAC unit proposed in~\cite{bitfusion} to accommodate the diverse bit-widths.
Fig.~\ref{fig:fu}-(a) illustrates how this MAC unit operates across multiple precision modes, i.e., 4-bit, 8-bit, and 16-bit. 
The unit consists of 16 multipliers that perform 4-bit by 4-bit multiplications and enables bit-level flexibility by dynamically fusing their outputs.

\begin{figure}[t]
    \centering
    \includegraphics[scale=0.67]{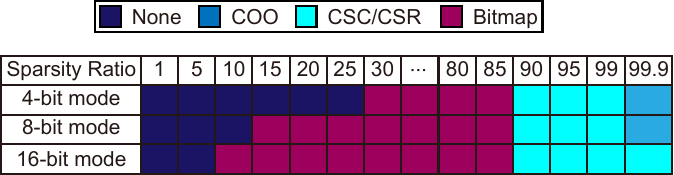}
    \caption{Optimal sparsity formats achieving the lowest memory footprint at different sparsity ratios in 4-bit, 8-bit, and 16-bit modes.}
    \label{fig:optimal_format}\vspace{-2mm}
\end{figure}

Fig.~\ref{fig:fu}-(b) shows the structure of a 2-dimensional (2D) 64 $\times$ 64 MAC array composed of the bit-scalable MAC units. 
This MAC array features varying data fetch sizes depending on the supported precision level. 
Specifically, as the precision decreases, the data fetch size doubles. 
This is because reducing the precision by half results in a fourfold increase in the number of multipliers, which are arranged in a 2D array structure.
Due to this characteristic, the ratio of data to metadata varies when using sparsity formats in the MAC array, and the sparsity ratio that minimizes the memory footprint changes depending on the precision and sparsity levels \textbf{\textit{(Takeway 4)}}. 
Fig.~\ref{fig:sparsity_ratio} presents an analysis of memory footprint across various sparsity ratios and precision levels with different compression formats, i.e., None, COO, CSC/CSR, and Bitmap. 
As shown, lower precision levels shift the graphs of sparsity formats to the right relative to uncompressed data (None), and the memory reduction effect becomes more significant, leading to an expansion of the y-axis.
Fig.~\ref{fig:optimal_format} summarizes the sparsity formats that achieve the lowest memory footprint for each supported precision mode across various sparsity ratios.
A larger memory footprint leads to increased memory accesses, which in turn degrade performance and increase energy consumption~\cite{eyeriss_v1, dram_cache}.
Therefore, a bit-scalable MAC array performing GEMM/GEMV operations must adaptively support sparsity formats based on precision and sparsity ratios \textbf{\textit{(Design requirement 3)}}.

\vspace{+4mm}
\section{Key Architectural Features of a GEMM/GEMV Unit in FlexNeRFer}\label{sec:architecture_feature}
This section presents the key architectural features of FlexNeRFer, which are designed to meet the design requirements essential for accelerating various NeRF models, as outlined in [Section~\ref{sec:gem_gemv_acceleration}].
These features include: i) a distribution network, ii) a reduction tree, and iii) an adaptive sparsity format support.
The distribution network and reduction tree enable multi-dataflow and sparsity support on the bit-scalable MAC array, making dense data mapping possible.
The adaptive sparsity format mechanism enhances data processing efficiency by realizing real-time dynamic data compression.

\begin{figure}[t]
    \centering
    \includegraphics[scale=0.53]{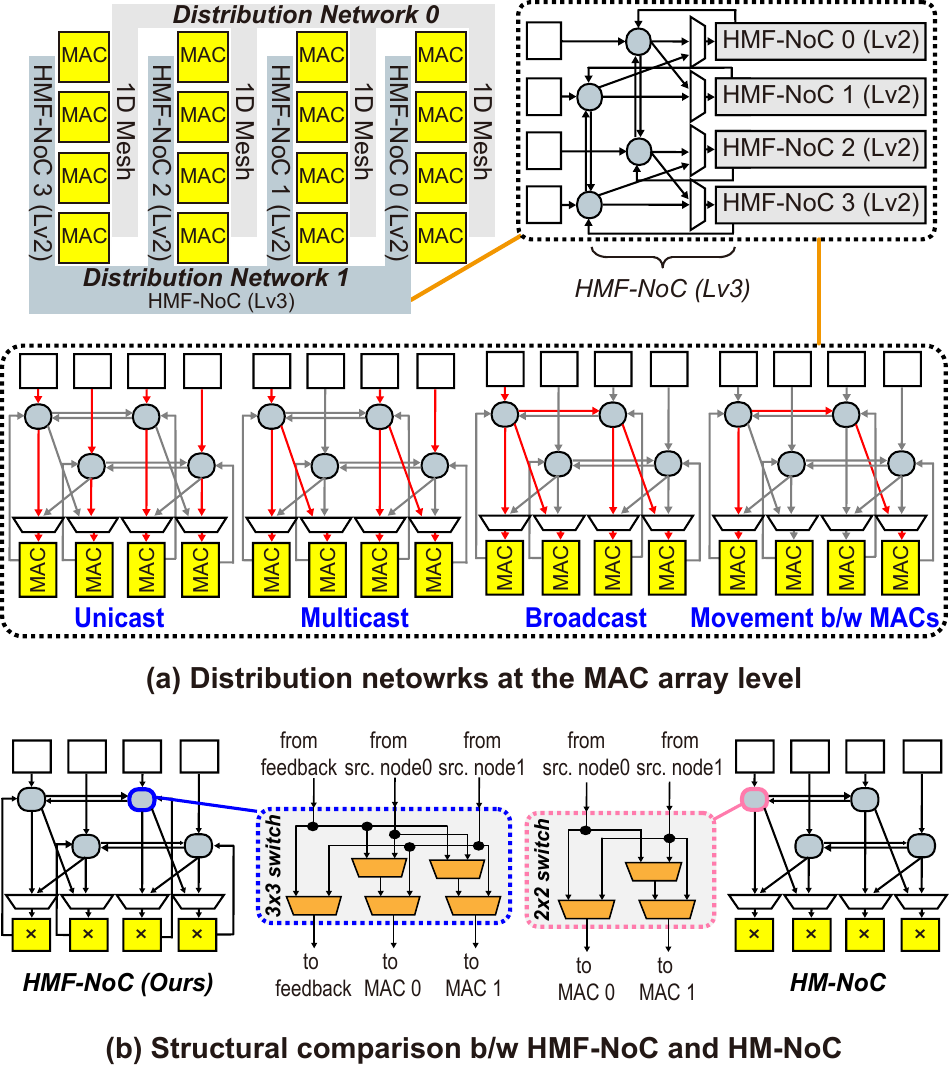}\vspace{-2mm}
    \caption{(a) Overview of the distribution network (DN) in FlexNeRFer’s MAC array. (b) Structural comparison of HMF-NoC (ours) and HM-NoC~\cite{eyeriss_v2}.}
    \label{fig:overall_noc}\vspace{-3mm}
\end{figure}

\subsection{Distribution Network (DN)}\label{sec:distribution_noc}

\subsubsection{High-Level Overview}\label{sec:overview_overview}
\textbf{\newline}As shown in Fig.~\ref{fig:fu}, the bit-scalable MAC array consists of MAC units at the array level and sub-multipliers within each MAC unit.
Since data can be sparsely mapped to both MAC units and their sub-multipliers, FlexNeRFer employs a hierarchical and flexible interconnect to ensure efficient dense data mapping to the MAC units and their sub-multipliers.
Specifically, at the array level, a hierarchical mesh with feedback network-on-chip (HMF-NoC) and a 1D mesh are utilized, while within each MAC unit, a lower-level HMF-NoC and column-level bypass links (CLBs) serve as the DN.

\begin{figure}[t]
    \centering
    \includegraphics[scale=0.67]{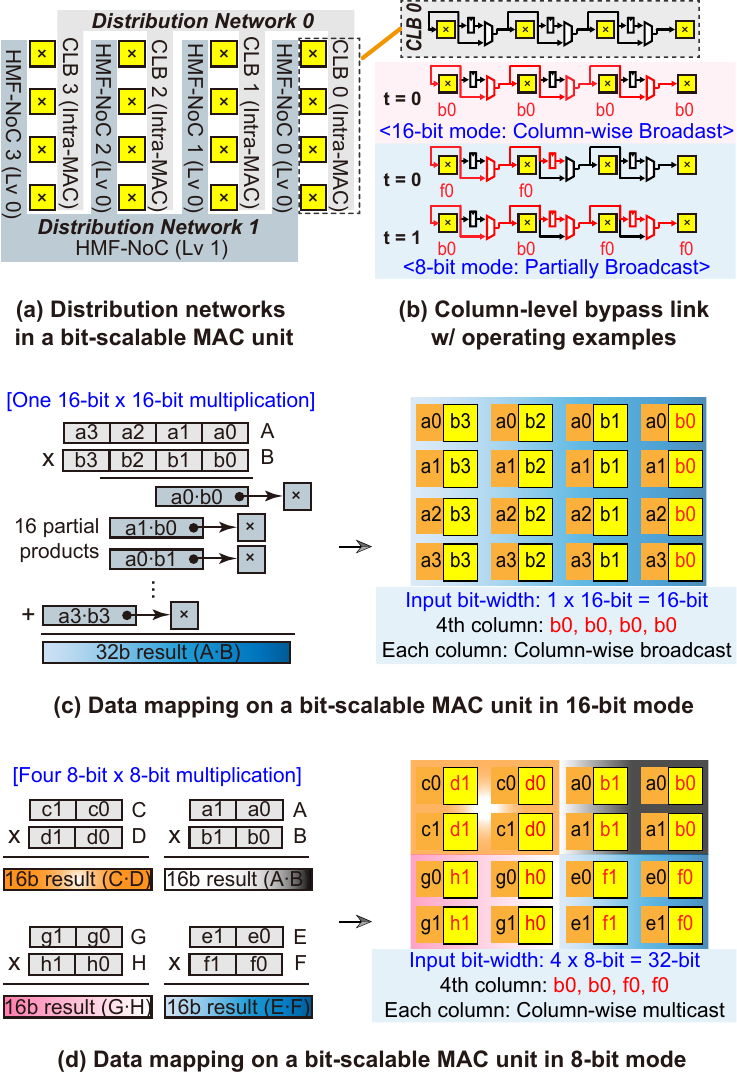}
    \caption{(a) Distribution network (DN) in a bit-scalable MAC unit. (b) Schematic diagrams illustrating the operations of the column-level bypass link (CLB) in 4- and 8-bit modes. (c-d) Data mapping in 16- and 8-bit modes on a bit-scalable MAC unit.}
    \label{fig:network_blc}\vspace{-4mm}
\end{figure}

\subsubsection{Array-Level NoC}\label{sec:overview_DN_Array}
\textbf{\newline}
As depicted in Fig.~\ref{fig:moti_dataflow}, the elements of one matrix are transmitted in a unicast manner, while those of the other matrix are delivered to either a single MAC unit or multiple MAC units.
To support the former, FlexNeRFer employs a 1D mesh NoC, whereas the latter benefits from efficient data mapping through a hierarchical mesh network with feedback (HMF-NoC) structure.
Fig.~\ref{fig:overall_noc} shows the overall DN in the MAC array of FlexNeRFer and the detailed architecture of HMF-NoC, which supports various dataflows, i.e., 1D broadcast, 1D multicast, and 1D unicast.
HMF-NoC is an extended NoC based on HM-NoC introduced in~\cite{eyeriss_v2}, incorporating a feedback loop, and changing each node from a 2 $\times$ 2 switch to a 3 $\times$ 3 switch, which facilitates data movement between MAC units and thereby reduces energy consumption for on-chip memory access.
To evaluate its impact, we modified an open-source cycle-level simulator, i.e., STONNE~\cite{stonne}, and modeling with the SRAM power-performance-area (PPA) tool, i.e., CACTI 6.0~\cite{cacti}, to assess the energy cost, finding that HMF-NoC consumes approximately 2.5$\times$ less energy for on-chip memory access compared to HM-NoC.
FlexNeRFer organizes HMF-NoC in a hierarchical structure, integrating both the MAC array and individual MAC units to efficiently distribute data across multiple levels.
The upper-level HMF-NoC (i.e., HMF-NoC (Lv3) in Fig.~\ref{fig:overall_noc}-(a)) delivers data to the columns of the array using various dataflow schemes, while the lower-level HMF-NoC (i.e., HMF-NoC (Lv2) in Fig.~\ref{fig:overall_noc}-(a)) distributes the data from the upper array level to the MAC units in each row.

\begin{figure*}[t]
    \centering
    \includegraphics[scale=0.54]{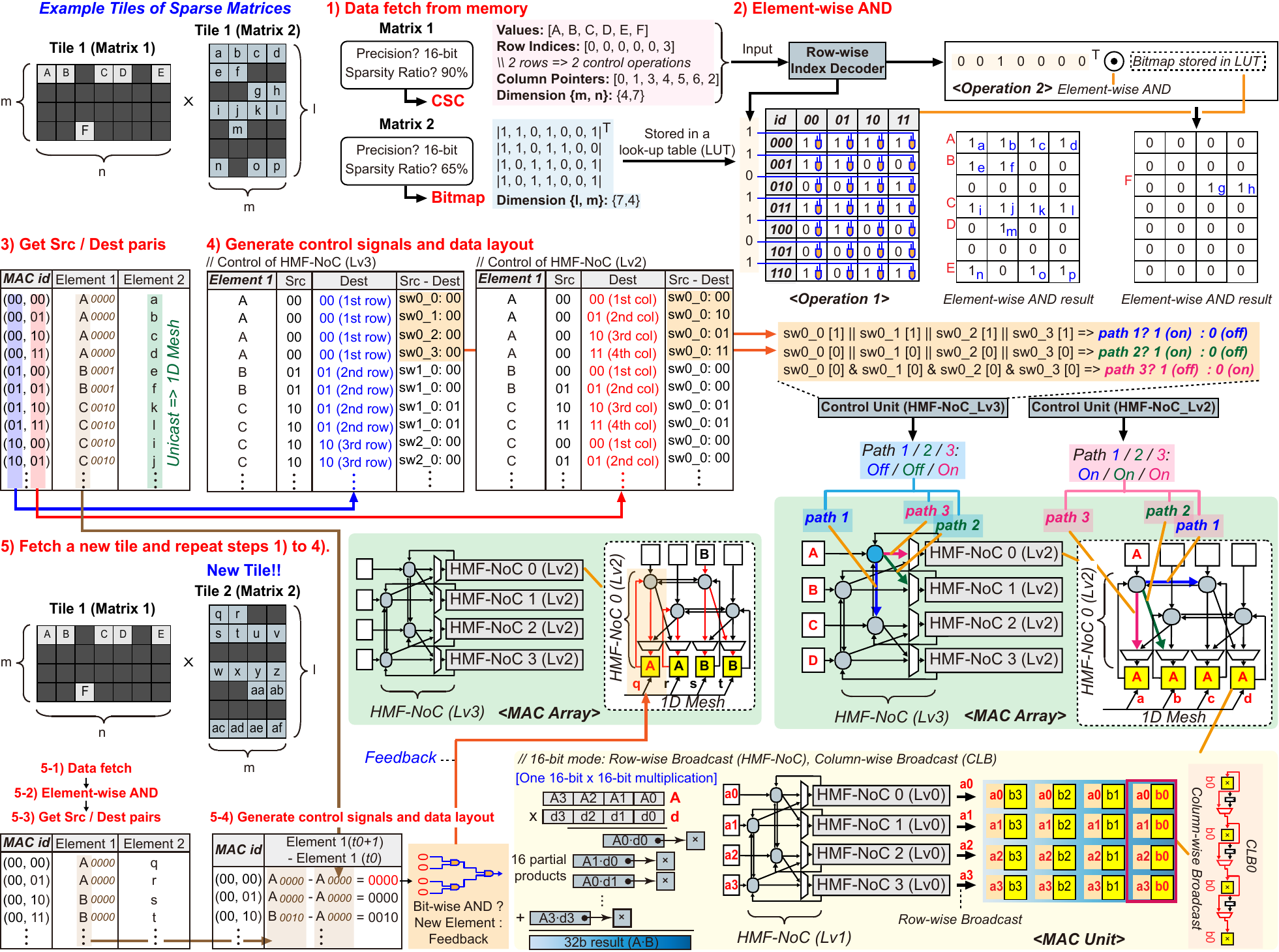}\vspace{-2mm}
    \caption{Walkthrough illustrating how data for sparse irregular GEMM operations is mapped onto a 16 $\times$ 16 MAC array in 16-bit mode.}\vspace{-2mm}
    \label{fig:overall_mac}
\end{figure*}

\subsubsection{MAC Unit-Level NoC}\label{sec:mac_level_DN_MAC}
\textbf{\newline}
As shown in Fig.~\ref{fig:fu}-(a), the bit-scalable MAC unit employed in FlexNeRFer consists of sixteen sub-multipliers. 
Similar to the array-level MAC units, these sub-multipliers can receive sparse data. 
To achieve efficient dense mapping, the same mechanism used at the array level is applied. 
Specifically, the elements of one matrix are transmitted in a unicast manner, while those of the other matrix are distributed across sub-multipliers in various dataflow patterns, such as unicast, multicast, and broadcast, along the column or row direction.
FlexNeRFer handles these cases by employing a column-level bypass link (CLB) for unicast transmission and a two-level HMF-NoC for multi-pattern data transfer (Fig.~\ref{fig:network_blc}-(a)). 

\begin{figure*}[ht]
    \centering
    \includegraphics[scale=0.71]{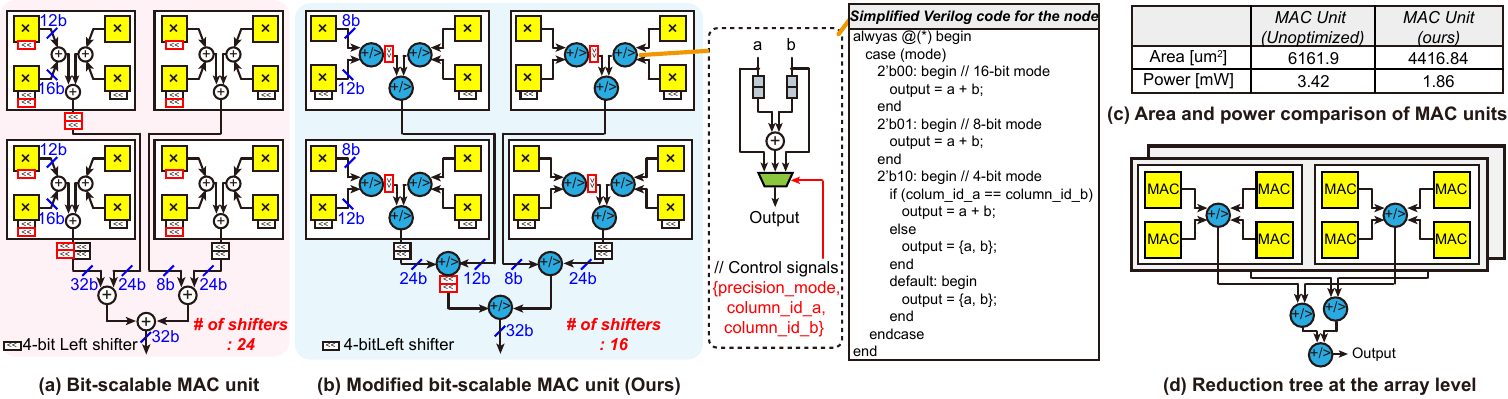}\vspace{-2mm}
    \caption{(a) Schematic diagram of the bit-scalable MAC unit from~\cite{bitfusion}. 
    (b) Reduction tree (RT) structure at the MAC unit level of FlexNeRFer.
    (c) Area and power comparison between the MAC unit with an optimized shifter-based RT (proposed) and the MAC unit without this optimization. 
    (d) RT architecture at the array level of FlexNeRFer.}
    \label{fig:accumulator}\vspace{-2mm}
\end{figure*}

Unlike the communication fabric at the MAC array level, the CLB is utilized for unicast transmission at the MAC unit level. 
Fig.~\ref{fig:network_blc}-(b) illustrates the detailed structure of the CLB, which utilizes pipelined paths to mitigate bandwidth (BW) variations across different precision modes in the bit-scalable MAC unit. 
More specifically, Fig.~\ref{fig:network_blc}-(c-d) show how data is mapped to a MAC unit in 16-bit and 8-bit modes, respectively. 
In 16-bit mode, the 16-bit input data of each operand is mapped to the MAC unit, while in 8-bit mode, the 32-bit input data is mapped. 
Although not depicted in the figure, in 4-bit mode, the 64-bit input data of each operand is mapped to the MAC unit.
Since the bandwidth is provisioned based on the 4-bit mode, the bit-scalable MAC unit is constrained by lower bandwidth (BW) utilization , i.e., 25\% (= 16b/64b) and 50\% (= 32b/64b) in the 16-bit and 8-bit modes, respectively.

\begin{figure*}[t]
    \centering
    \includegraphics[scale=0.68]{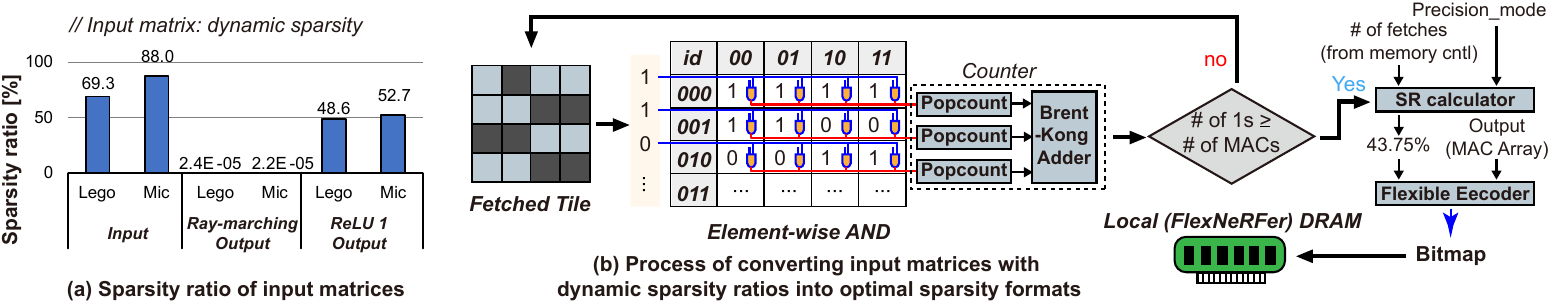}
    \caption{(a) Sparsity ratios of matrices at different stages of the rendering process in the Instant-NGP model~\cite{instant_ngp}, for two scenes from the Synthetic-NeRF dataset~\cite{synthetic_nerf}. 
    (b) Sparsity-aware data compression process applied to input data in FlexNeRFer.}\label{fig:adaptive_sparsity_format}\vspace{-2mm}
\end{figure*}

The CLB, consisting of 16 wired links, transmits input data in 16-bit units to overcome the low BW utilization issue, ensuring 100\% BW utilization regardless of the precision mode.
Furthermore, the CLB supports a precision-adaptive forwarding scheme through bypassable wired links. 
In higher precision modes (8-bit and 16-bit), subwords of input data need to be delivered to multiple sub-multipliers. 
The CLB facilitates this process by leveraging bypassable links, ensuring that the required data can be transmitted to the sub-multipliers within a single cycle.
Fig.~\ref{fig:network_blc}-(c-d) and Fig.~\ref{fig:network_blc}-(b) show the operation of the CLB by depicting how data are mapped onto the MAC array in 8-bit and 16-bit modes and how the CLB operates in these modes.
In 16-bit mode, the identical subword of an operand is mapped to all sub-multipliers within each column of the MAC unit (e.g., the sub-data ‘b0’ in the 4th column, as shown in Fig.~\ref{fig:network_blc}-(c)). 
In 8-bit mode, the same data is mapped to the sub-multipliers in the first and second rows, as well as the third and fourth rows (e.g., the sub-data ‘b0’ and ‘f0’ in the 4th column, as shown in Fig.~\ref{fig:network_blc}-(d)).
In such cases, the bypassable links forward data between sub-multipliers using broadcast and multicast approaches, respectively, allowing identical data to be delivered to all relevant sub-multipliers with a single data fetch (Fig.~\ref{fig:network_blc}-(b)).

\vspace{+1mm}\noindent\textbf{{Mapping Process Example:}} Fig.~\ref{fig:overall_mac} shows how the DN of FlexNeRFer operates on a 16 $\times$ 16 MAC array in 16-bit mode.


\subsection{Reduction Tree (RT)}\label{sec:reduction_noc}

The bit-scalable MAC unit supports high-precision multiplication by performing shift-addition operations on the results of internal sub-multipliers. 
Fig.~\ref{fig:accumulator}-(a) illustrates the schematic diagram of the bit-scalable MAC unit, which includes shifters and a reduction tree (RT) to support the operations.
A MAC array built with this unit requires a large number of shifters, e.g., 6,144 shifters for a 16$\times$16 MAC array.
To reduce this overhead, FlexNeRFer optimizes the MAC unit by sharing shifters that perform identical operations, as shown in Fig.~\ref{fig:accumulator}-(b).
With this design improvement, the MAC unit of FlexNeRFer reduces the number of shifters by 33.3\% (= 24$\rightarrow$16) compared to the bit-scalable MAC unit without shifter optimization, thereby reducing the overall number of shifters at the MAC array level.
In addition to shifter optimization, the RT of FlexNeRFer incorporates a comparator and a bypassable adder, similar to the works of~\cite{flexagon, trapezoid, gamma}.
They enable flexible reduction by either adding two operands if their indices of sparsity formats match, or directly passing them through if they do not.
Fig.~\ref{fig:accumulator}-(c) presents the area and power comparison between the MAC unit supporting flexible reduction with optimized shifters and the non-optimized shifters. 
The optimized MAC unit achieves an area reduction of 28.3\% and a power reduction of 45.6\% compared to the non-optimized counterpart.
Fig.~\ref{fig:accumulator}-(d) shows the RT architecture at the MAC array level. 
FlexNeRFer employs a flexible augmented reduction tree (ART) between MAC units, building upon the insights provided in the works of~\cite{flexagon, feather, maeri} which validated the area efficiency of this structure.

\subsection{Sparsity-Aware Data Compression}\label{sec:data_compression}
In Section~\ref{sec:need_adaptive_sparsity}, we observed that the optimal sparsity format, which minimizes memory footprint, varies depending on the sparsity ratio and the precision mode. 
To maximize data processing efficiency, FlexNeRFer compresses input and weight data into the optimal sparsity format and stores it in memory. 
Fig.~\ref{fig:adaptive_sparsity_format}-(a) shows the sparsity ratio of input data measured at various stages of the rendering process for two scenes from the Synthetic-NeRF dataset~\cite{synthetic_nerf} using the Instant-NGP model~\cite{instant_ngp}.
The sparsity ratio varies across rendering stages, indicating that FlexNeRFer needs to identify the sparsity ratio in real-time during the rendering process to support the optimal sparsity format.
Fig.~\ref{fig:adaptive_sparsity_format}-(b) illustrates this process, showing how FlexNeRFer identifies the sparsity ratio of the input data and converts it into the optimal format.
The sparsity ratio ($SR$) in FlexNeRFer is calculated by
\begin{equation}\vspace{-1mm}
SR (\%) = \left( 1 - \frac{\sum_{i=1}^{N_{\text{fetch}}} \text{Popcount}(\text{Fetched Tile}_i)}{N_{\text{fetch}} \times N_{\text{data/fetch}}} \right) \times 100,
\end{equation}
where $N_{\text{fetch}}$ represents the number of data fetches required to fully map the data onto the MAC array, and $N_{\text{data/fetch}}$ denotes the number of elements per fetch.
As described in Section.~\ref{sec:need_adaptive_sparsity}, $N_{\text{data/fetch}}$ increases fourfold when the data precision is halved, due to the doubling of the fetch size.
Based on the sparsity ratio and precision mode, FlexNeRFer compresses data by adaptively selecting the optimal sparsity format through a flexible format encoder.
For weight data, since weights remain fixed after training, FlexNeRFer precomputes the sparsity ratio and converts them into the optimal sparsity format before storing them in its local DRAM.

\section{Overall Architecture of FlexNeRFer}\label{sec:flexnerfer_archi}

\subsection{Architecture Overview}\label{sec:architectural_overview}
Fig.~\ref{fig:overall_Arcthiecture} presents the overall architecture of FlexNeRFer.
FlexNeRFer consists of a RISC-V controller, which decodes programs copied from the host and generates global control signals, a DMA engine for efficient data transfer between host memory and local memory, local DRAM for storing data copied from the host, a NeRF encoding unit that supports the encoding processes of NeRF, and a GEMM/GEMV acceleration unit.
The GEMM/GEMV acceleration unit includes the interconnect, bit-scalable MAC array, and flexible format encoder/decoder discussed in the previous section (i.e., Section~\ref{sec:architecture_feature}), which are designed to support sparsity and quantization.

\subsection{NeRF Encoding Unit}\label{sec:nerf_encoding_unit}
The neural feature encoding process is a major bottleneck in various NeRF models.
FlexNeRFer addresses this challenge with specialized positional and hash encoding modules.

\subsubsection{Positional Encoding Engine (PEE)}
\textbf{\newline}
As shown in Eq.~(\ref{eq:postional_encoding}) in Section~\ref{sec:nerf_concept}, the positional encoding transforms input coordinates into a high-dimensional feature space using trigonometric functions.
According to~\cite{metavrain_journal}, trigonometric function values can be approximated using the following equations without any degradation in image quality through fine-tuning:
\begin{equation}
    \sin(2^{-1} \pi v) \approx (-1)^{\lfloor v/2 \rfloor} \cdot \text{mod}(v, 2) \cdot \text{mod}(2 - v, 2),
    \label{eq:approximation_sin}
\end{equation}
\begin{equation}
    \cos(2^{-1} \pi v) \approx (-1)^{\lfloor v/2 \rfloor} \cdot \text{mod}(v+1, 2) \cdot \text{mod}(1 - v, 2).
    \label{eq:approximation_cos}
\end{equation}
Leveraging this approximation, the positional encoding engine (PEE) in FlexNeRFer employs dedicated processing units for efficient trigonometric computation.
In Eq.~(\ref{eq:approximation_sin}) and (\ref{eq:approximation_cos}), the modulo operation is implemented using an arithmetic bit-shifter.
This implementation enables the PEE to processes 64 positional encodings in parallel, achieving an 8.2$\times$ area reduction and a 12.8$\times$ power reduction compared to a DesignWare IP-based PEE from Synopsys~\cite{dw_synopsys}.

\begin{figure}[t]
    \centering
    \includegraphics[scale=0.73]{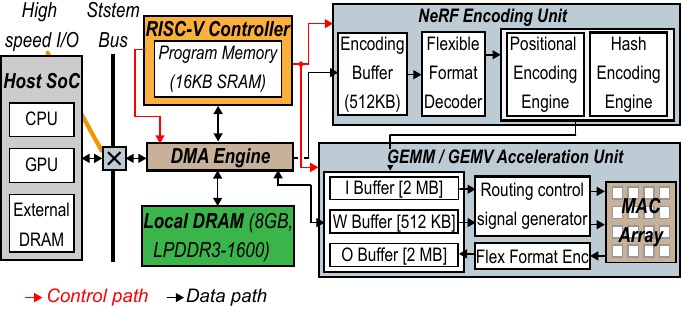}
    \caption{Overall architecture of the FlexNeRFer accelerator.}\label{fig:overall_Arcthiecture}\vspace{-2mm}
\end{figure}

\subsubsection{Hash Encoding Engine (HEE)}
\textbf{\newline} 
Various NeRF models use hash encoding to efficiently represent spatial information via a multi-resolution hash table, avoiding costly trigonometric computations. 
This process involves i) hash table lookup and ii) interpolation to generate neural network input features.
A key challenge in multi-resolution hash table lookup is balancing memory efficiency and accuracy. At low resolutions, multiple coordinates share the same hash index, causing frequent hash collisions, whereas at high-resolution levels, large hash table sizes exceed on-chip memory capacity limits, leading to increased DRAM accesses~\cite{neurex, neu_gpu, instant_3d}.
To address this, the Hash Encoding Engine (HEE) is built upon and extended from the hardware unit proposed in NeuRex~\cite{neurex}.
The HEE in FlexNeRFer incorporates 64 coalescing hash units, which reduce redundant memory accesses by grouping data with the same hash index into a single block and processing them in a coalesced manner at low-resolution levels. 
It also employs 64 subgrid hash units, which divide the grid into sub-grids and encode data using smaller hash tables at high-resolution levels. Additionally, it includes 64 interpolation units that support parallel trilinear interpolation computation.

\begin{table}[t]
\centering
\caption{Hardware specifications comparison between the MAC array of FlexNeRFer and multiple baselines for GEMM/GEMV operations.} 
\label{tab:design_spec}\vspace{-1mm}
\scalebox{0.72}{
\begin{tabular}{|c|cccc|}
\hline
\textbf{Compute Array}                                                               & \multicolumn{1}{c|}{\textbf{SIGMA}} & \multicolumn{1}{c|}{\textbf{Bit Fusion}}  & \multicolumn{1}{c|}{\textbf{\begin{tabular}[c]{@{}c@{}}Bit-Scalable\\ SIGMA\end{tabular}}} & \textbf{\begin{tabular}[c]{@{}c@{}}MAC Array\\ (FlexNeRFer)\end{tabular}} \\ \hline\hline
\textbf{Technology}                                                                  & \multicolumn{4}{c|}{Commercial CMOS 28nm}                                                                                                                                                                                                                \\ \hline
\textbf{Clock Frequency}                                                             & \multicolumn{4}{c|}{800 MHz}                                                                                                                                                                                                                             \\ \hline
\textbf{Bit-flexibility?}                                                            & \multicolumn{1}{c|}{\xmark}             & \multicolumn{1}{c|}{\cmark}                  & \multicolumn{1}{c|}{\cmark}                                                                   & \cmark                                                                       \\ \hline
\textbf{Sparsity?}                                                                   & \multicolumn{1}{c|}{\cmark}            & \multicolumn{1}{c|}{\xmark}                   & \multicolumn{1}{c|}{\cmark}                                                                   & \cmark                                                                       \\ \hline
\textbf{\begin{tabular}[c]{@{}c@{}}Supported\\ Precision\end{tabular}}               & \multicolumn{1}{c|}{INT16}          & \multicolumn{1}{c|}{INT (4 / 8 / 16)} & \multicolumn{1}{c|}{INT (4 / 8 / 16)}                                                  & INT (4 / 8 / 16)                                                      \\ \hline
\textbf{\# of multipliers}                                                           & \multicolumn{1}{c|}{$64^2$}             & \multicolumn{1}{c|}{$64^2$ /$128^2$ / 256}       & \multicolumn{1}{c|}{$64^2$ / $128^2$ / $256^2$}                                                        & $64^2$ / $128^2$ / $256^2$                                                            \\ \hline
\textbf{Area {[}mm$^2${]}}                                                                & \multicolumn{1}{c|}{20.5}           & \multicolumn{1}{c|}{31.9}                 & \multicolumn{1}{c|}{40.8}                                                                  & 28.6                                                                      \\ \hline
\textbf{Power {[}W{]}}                                                              & \multicolumn{1}{c|}{5.8}            & \multicolumn{1}{c|}{5.8 / 5.3 / 4.8}      & \multicolumn{1}{c|}{9.3 / 8.7 / 8.2}                                                       & 6.9 / 6.4 / 5.5                                                           \\ \hline
\textbf{\begin{tabular}[c]{@{}c@{}}Peak Efficiency\\ {[}TOPS/W{]}\end{tabular}}      & \multicolumn{1}{c|}{1.1}            & \multicolumn{1}{c|}{18.1 / 4.9 / 1.4}     & \multicolumn{1}{c|}{5.7 / 3.0 / 0.8}                                                       & 15.2 / 4.1 / 1.2                                                          \\ \hline
\textbf{\begin{tabular}[c]{@{}c@{}}Effective Efficiency\\ {[}TOPS/W{]}\end{tabular}} & \multicolumn{1}{c|}{1.0}            & \multicolumn{1}{c|}{3.2 / 0.8 / 0.2}      & \multicolumn{1}{c|}{4.4 / 2.5 / 0.7}                                                       & 11.8 / 3.4 / 1.2                                                          \\ \hline
\end{tabular}
}\vspace{2mm}
\end{table}

\begin{figure}[t]
    \centering
    \includegraphics[scale=0.69]{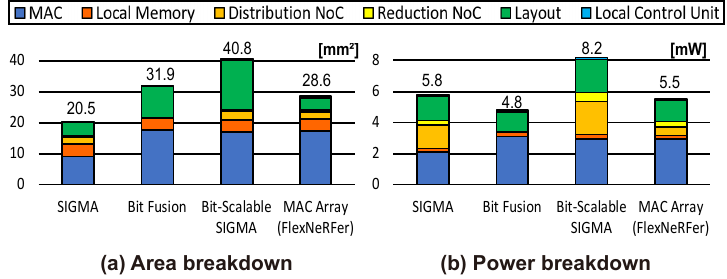}\vspace{-2mm}
    \caption{(a) Area and (b) power breakdowns of compute arrays supporting GEMM/GEMV operations. Power consumption is reported for the INT16 precision mode.}\label{fig:mac_array_comparison}\vspace{-1mm}
\end{figure}

\begin{figure}[t]
    \centering
    \includegraphics[scale=0.80]{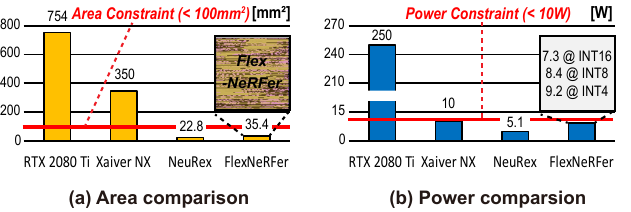}\vspace{-2mm}
    \caption{(a) Area and (b) power comparison of FlexNeRFer with GPUs (NVIDIA RTX 2080 Ti and NVIDIA Xavier NX) and a NeRF accelerator (NeuRex).}\label{fig:area_comparison}\vspace{-2mm}
\end{figure}

\section{Evaluation}\label{sec:evaluation}

\subsection{Methodology}\label{sec:baselines}
\textbf{Baselines: }To access the effectiveness of the network topology employed in FlexNeRFer, we compared hardware cost across various NeRF models against three GEMM/GEMV processing arrays: SIGMA~\cite{sigma}, Bit Fusion~\cite{bitfusion}, and bit-scalable SIGMA~\cite{sigma, bitfusion}. SIGMA is a processing array that supports sparsity through Benes network and FAN topology but does not support bit-scalability, whereas Bit Fusion supports multiple data formats, including INT4, INT8, and INT16, but lacks sparsity support. 
Bit-scalable SIGMA integrates the NoC structure of SIGMA into the MAC array of Bit Fusion.
In addition, we compared FlexNeRFer with a recent NeRF accelerator, i.e., NeuRex~\cite{neurex} which supports INT16, and NVIDIA RTX 2080 Ti~\cite{rtx_2080_ti} to evaluate the efficiency in the NeRF rendering process.


\noindent\textbf{Workload and Dataset: }We evaluated seven representative NeRF models—NeRF~\cite{vanilar_nerf}, KiloNeRF~\cite{kilonerf}, NSVF~\cite{NSVF}, Mip-NeRF~\cite{mip_nerf}, Instant-NGP~\cite{instant_ngp}, IBRNet~\cite{ibrnet}, and TensoRF~\cite{TensoRF}—on two widely used datasets, i.e., NeRF-Synthetic~\cite{vanilar_nerf} and NSVF~\cite{NSVF}, both with an image resolution of 800$\times$800 and a batch size of 4096.

\begin{figure}[t]
    \centering
    \includegraphics[scale=0.76]{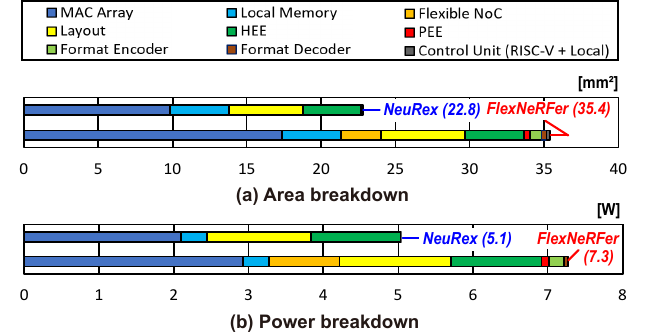}\vspace{-2mm}
    \caption{(a) Area and (b) power breakdowns of FlexNeRFer and the baseline NeRF accelerator, i.e., NeuRex. For FlexNeRFer, power consumption at INT16 mode is reported.}\label{fig:breakdown_nerf_acc}
\end{figure}

\begin{figure}[t]
    \centering
    \includegraphics[scale=0.77]{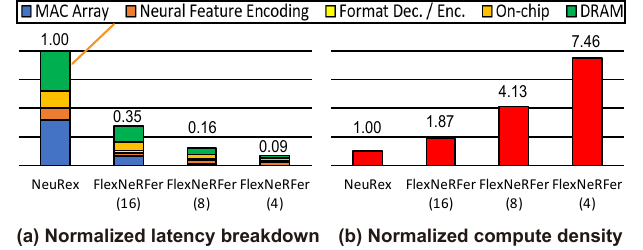}\vspace{-2mm}
    \caption{Comparison of normalized latency and compute density, i.e., area efficiency, between NeuRex and FlexNeRFer. 
    Numbers in parentheses for FlexNeRFer indicate the operating precision mode}.\label{fig:acc_latency_area_efficiency}\vspace{-3mm}
\end{figure}

\begin{figure*}[t]
    \centering
    \includegraphics[scale=0.72]{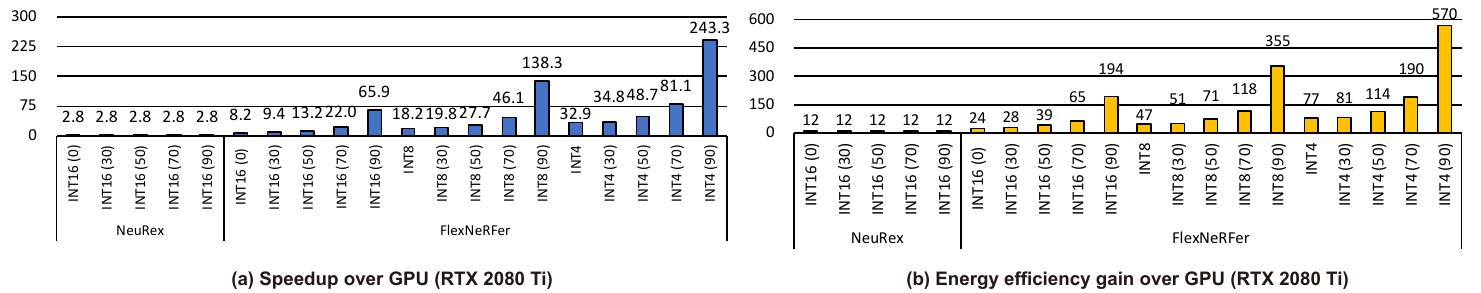}\vspace{-2mm}
    \caption{(a) Speedup and (b) Energy efficiency gain over NVIDIA RTX 2080 Ti. Numbers in parentheses indicate the pruning ratio.}\label{fig:archi_evaluation}\vspace{-1mm}
\end{figure*}

\noindent{\textbf{Design Implementation:}} To obtain both netlist and post-routing hardware costs, we performed synthesis and place-and-route (PnR) using Synopsys EDA tools with a 28nm CMOS process under fast/fast and slow/slow corners. 
The area information was extracted by using Synopsys IC Compiler~\cite{synopsys_icc}. 
Additionally, we report power consumption estimated by Synopsys PrimeTime PX~\cite{pt_synopsys} using SAIF (Switching Activity Interchange Format) data from post-layout simulation, along with parasitic data from Synopsys StarRC~\cite{rc_synopsys} and PnR results.
For cycle estimation, we modified the open-source cycle-level simulator STONNE~\cite{stonne} to reflect the dataflow and memory configurations of FlexNeRFer and the baseline architectures when estimating the number of compute cycles (= latency).

\noindent{\textbf{Memory Modeling:}}
For system-level analysis, we considered both on-chip and off-chip memory accesses when evaluating performance and energy consumption. 
For on-chip memory, power and timing information were obtained from a memory compiler, while for off-chip memory, we used the timing and power specifications of LPDDR3~\cite{lpddr3} (Fig.~\ref{fig:overall_Arcthiecture}). 

\subsection{GEMM/GEMV Acceleration}\label{sec:Design_implementation}
Table~\ref{tab:design_spec} presents a comparison of the design specifications of the MAC array in FlexNeRFer and the baseline compute MAC arrays, and Fig.~\ref{fig:mac_array_comparison} shows the area and power breakdowns of the arrays.
FlexNeRFer occupies 1.4$\times$ larger area than SIGMA, which results from the adoption of a large bit-scalable MAC array to support bit-flexibility, but it consumes 5.2\% less power in INT16 mode due to its power-efficient interconnect structure resulting from a reduced number of switching nodes [Section~\ref{sec:overview_DN_Array}].
Compared to Bit Fusion, FlexNeRFer achieves area reduction by a 10.3\%, which is due to the pipeline structure applied to the datapath, reducing the layout area of the datapath [Section~\ref{sec:mac_level_DN_MAC}].
Bit-scalable SIGMA employs a flexible NoC with a greater number of switching nodes [Section~\ref{sec:overview_DN_Array}] and unoptimized shifters [Section~\ref{sec:reduction_noc}], leading to an array area that is 1.4$\times$ larger than FlexNeRFer, while consuming 1.3$\sim$1.5$\times$ more power across the supported precision modes.
Meanwhile, the MAC array in FlexNeRFer achieves effective efficiency that is 1.2$\sim$11.8$\times$ higher than the baselines, owing to its power-efficient NoC design, support for quantization via a bit-scalable MAC array, and dense data mapping that eliminates sparse data, leading to improved computational resource utilization.


\vspace{-2mm}
\subsection{Rendering Acceleration}\label{sec:gemm_comparison}

\subsubsection{Hardware Cost Analysis}\label{sec:hardware_cost_analysis}
\textbf{\newline}
Fig.~\ref{fig:area_comparison} presents the area and power consumption of FlexNeRFer, a baseline NeRF accelerator, i.e., NeuRex, and GPU devices.
NeuRex and FlexNeRFer occupy areas of 22.8 mm$^{2}$ and 35.4 mm$^{2}$, respectively, with power consumptions of 5.1 W and 7.3$\sim$9.2 W.
As mentioned in Section~\ref{sec:intro}, an on-device accelerator is generally expected to have an area smaller than 100 mm$^{2}$ and a power consumption below 10 W to be viable for integration.
Both NeuRex and FlexNeRFer meet these constraints.
Fig.~\ref{fig:breakdown_nerf_acc} provides a detailed breakdown of the area and power in FlexNeRFer and NeuRex.
Unlike NeuRex, FlexNeRFer integrates a precision-scalable MAC array to enable bit-scalability and a flexible NoC to bypass computations on sparse data.
These features result in area overhead of 48.4\% and power overhead of 35.18\% in FlexNeRFer compared to NeuRex with significant reduction in compute latency.
Additionally, FlexNeRFer supports real-time format encoding and decoding for data fetched from or loaded into memory.
This format conversion process incurs a 3.2\% area overhead and a 3.4\% increase in power consumption.
As shown in Fig.~\ref{fig:acc_latency_area_efficiency}-(a), FlexNeRFer spends 8.7\% of its total execution time on format conversion in 16-bit mode.
However, this process reduces memory accesses, cutting DRAM access time by 72\%, while the flexible NoC accelerates MAC array computation by 4.6$\times$, ultimately reducing total execution time by 65\%.
Furthermore, this latency reduction becomes even more significant at lower precision modes, as illustrated in Fig.~\ref{fig:acc_latency_area_efficiency}-(a).
With this advantage, FlexNeRFer achieves higher compute density than NeuRex, despite having a larger area.
Fig.~\ref{fig:acc_latency_area_efficiency}-(b) shows that FlexNeRFer achieves 1.9$\sim$7.5$\times$ higher performance per unit area.
Finally, Fig.~\ref{fig:archi_evaluation} illustrates the improvements in speed and energy efficiency of NeuRex and FlexNeRFer compared to GPUs when structured pruning is applied.
Since NeuRex does not support sparsity or precision flexibility, its speedup and energy efficiency remain constant regardless of structured pruning. 
In contrast, FlexNeRFer demonstrates increasing performance and efficiency as structured pruning is applied.
Specifically, FlexNeRFer achieves 8.2$\sim$243.3$\times$ speedup and 24.1$\sim$520.3$\times$ improvement in energy efficiency over a GPU, while demonstrating 4.2$\sim$86.9$\times$ speedup and 2.3$\sim$47.5$\times$ improvement in energy efficiency compared to NeuRex.

\begin{figure}[t]
    \centering
    \includegraphics[scale=0.95]{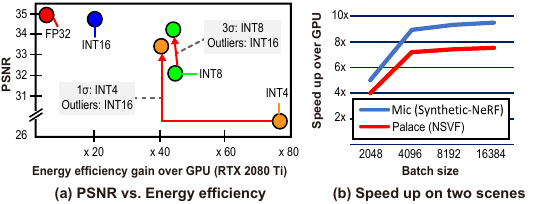}
    \caption{(a) PSNR vs. energy efficiency gain over NVIDIA RTX 2080 Ti at different precision modes. (b) Speedup over GPU when rendering scenes at different complexities and batch sizes.}
\label{fig:isca25_psnr}\vspace{-2mm}
\end{figure}

\subsubsection{Sensitivity Analysis}\label{sec:sensitivity_analysis}
\textbf{\newline}
\textbf{Impact of Quantization on PSNR: } To evaluate the impact of quantization on image quality, we quantized Instant-NGP~\cite{instant_ngp} to various bit-widths supported by FlexNeRFer and measured PSNR on the Synthetic-NeRF dataset~\cite{synthetic_nerf}. 
Fig~\ref{fig:isca25_psnr}-(a) presents the PSNR of the quantized NeRF models alongside their energy efficiency gains relative to a GPU.
For INT16, the PSNR remains nearly identical to the FP32 model (< 0.3 dB). 
However, INT4 and INT8 exhibit noticeable PSNR degradation (> 3 dB). 
To mitigate this while preserving the energy efficiency benefits of quantization, we applied a technique similar to \cite{first_ptq, int_only_ptq}, where a subset of data (i.e., outlier data) is represented in INT16. 
This method significantly improves image quality, with INT8 achieving near-FP32 PSNR and INT4 showing a difference in PSNR of less than 1.4 dB.

\noindent\textbf{Impact of Scene Complexity and Batch Size on Speedup:}
To analyze the impact of scene complexity and batch size on performance, we evaluated the rendering speed variations across different batch sizes for a \textit{simple} scene (Mic from NeRF-Synthetic~\cite{synthetic_nerf}) and a \textit{complex} scene (Palace from NSVF~\cite{NSVF}).
Our results show that the simple scene renders, on average, 1.2$\times$ faster than the complex scene, primarily due to fewer sampling points (Fig~\ref{fig:isca25_psnr}-(b)). 
Additionally, when the batch size exceeds 8192, performance gains plateau due to off-chip bandwidth limitations and insufficient computing resources.


\section{Conclusion}
In this work, we presented FlexNeRFer, an accelerator designed for fast and efficient rendering of diverse NeRF models on-device with low hardware cost.
It features a flexible NoC supporting multi-dataflow and sparsity on a precision-scalable MAC array, along with efficient data storage optimized for various sparsity and precision modes. 
Our evaluation demonstrates that FlexNeRFer achieves a speedup of 8.2$\sim$243.3$\times$ and an energy efficiency improvement of 24.1$\sim$520.3$\times$ compared to a GPU (NVIDIA RTX 2080 Ti). Additionally, it delivers a speedup of 4.2$\sim$86.9$\times$ and an energy efficiency gain of 2.3$\sim$47.5$\times$ over the state-of-the-art NeRF accelerator (NeuRex).
These results highlight the effectiveness of the architectural innovations of FlexNeRFer in enabling real-time, energy-efficient NeRF rendering for on-device applications.
We expect that FlexNeRFer can be effectively utilized in a wide range of on-device applications, including AR, VR, and autonomous driving systems.




\section{Acknowledgment}
This work was partially supported by the Institute of Information \& Communications Technology Planning \& Evaluation~(IITP) grant funded by the Ministry of Science and ICT (MSIT) under Grant No.2022-0-00991 and RS-2023-00229849, by the National Research Foundation of Korea~(NRF) funded by the Ministry of Science and ICT under Grant RS-2023-00258227, and 
by the Technology Innovation Program (RS-2024-00445759, Development of Navigation Technology Utilizing Visual Information Based on Vision-Language Models for Understanding Dynamic Environments in Non-Learned Spaces) funded by the Ministry of Trade, Industry \& Energy (MOTIE, Korea).
The EDA tool used in this work was supported by the IC Design Education Center (IDEC) in South Korea.

\bibliographystyle{plain}
\bibliography{ref}

\end{document}